%% file: main.tex
\documentclass[acmsmall,screen]{acmart}

\renewcommand\footnotetextcopyrightpermission[1]{} 
\usepackage{enumitem}
\usepackage{pifont}
\usepackage{multirow,makecell}
\usepackage{color,xcolor}
\usepackage{listings,amsfonts}
\usepackage{caption}
\usepackage{subcaption}
\usepackage{threeparttable}
\usepackage{bbding}
\usepackage{booktabs} 
\usepackage{longtable}
\usepackage{flushend}

\AtEndPreamble{
	\usepackage{hyperref}

}

\begin{document}

\title{Towards Reliable Vector Database Management Systems: A Software Testing Roadmap for 2030}

\author[S Wang]{Shenao Wang}
\email{shenaowang@hust.edu.cn}
\orcid{0000-0003-3818-3343}
\affiliation{%
  \institution{Huazhong University of Science and Technology}
  \city{Wuhan}           
  \country{China}
}

\author[Y Zhao]{Yanjie Zhao}
\email{yanjie_zhao@hust.edu.cn}
\orcid{0000-0001-8793-5367}
\affiliation{%
  \institution{Huazhong University of Science and Technology}
  \city{Wuhan}           
  \country{China}
}

\author[Y Xie]{Yinglin Xie}
\email{xieyinglin@hust.edu.cn}
\orcid{0009-0003-6153-0638}
\affiliation{%
  \institution{Huazhong University of Science and Technology}
  \city{Wuhan}           
  \country{China}
}

\author[Z Liu]{Zhao Liu}
\email{r3pwnx@gmail.com}
\orcid{0009-0002-7535-0227}
\affiliation{%
  \institution{360 AI Security Lab}
  \city{Beijing}           
  \country{China}
}

\author[X Hou]{Xinyi Hou}
\email{xinyihou@hust.edu.cn}
\orcid{0009-0005-9965-2109}
\affiliation{%
  \institution{Huazhong University of Science and Technology}
  \city{Wuhan}           
  \country{China}
}

\author[Q Zou]{Quanchen Zou}
\email{zouquanchen@gmail.com}
\orcid{0009-0004-5927-9680}
\affiliation{%
  \institution{360 AI Security Lab}
  \city{Beijing}           
  \country{China}
}

\author[H Wang]{Haoyu Wang}
\email{haoyuwang@hust.edu.cn}
\orcid{0000-0003-1100-8633}
\affiliation{%
  \institution{Huazhong University of Science and Technology}
  \city{Wuhan}           
  \country{China}
}

\input{Chapters/0.abstract}

\maketitle

\input{Chapters/1.introduction}
\input{Chapters/2.vdbms}
\input{Chapters/3.defects}
\input{Chapters/4.challenges}
\input{Chapters/5.conclusion}

\bibliographystyle{ACM-Reference-Format}
\bibliography{ref}

\end{document}

%% file: Chapters/0.abstract.tex
\begin{abstract}
The rapid growth of Large Language Models (LLMs) and AI-driven applications has propelled Vector Database Management Systems (VDBMSs) into the spotlight as a critical infrastructure component. VDBMS specializes in storing, indexing, and querying dense vector embeddings, enabling advanced LLM capabilities such as retrieval-augmented generation, long-term memory, and caching mechanisms. However, the explosive adoption of VDBMS has outpaced the development of rigorous software testing methodologies tailored for these emerging systems. Unlike traditional databases optimized for structured data, VDBMS face unique testing challenges stemming from the high-dimensional nature of vector data, the fuzzy semantics in vector search, and the need to support dynamic data scaling and hybrid query processing. In this paper, we begin by conducting an empirical study of VDBMS defects and identify key challenges in test input generation, oracle definition, and test evaluation. Drawing from these insights, we propose the first comprehensive research roadmap for developing effective testing methodologies tailored to VDBMS. By addressing these challenges, the software testing community can contribute to the development of more reliable and trustworthy VDBMS, enabling the full potential of LLMs and data-intensive AI applications.
\end{abstract}

%% file: Chapters/1.introduction.tex
\section{Introduction}
The advent of \textbf{Large Language Models (LLMs)} and the AI revolution has catapulted \textbf{Vector Database Management Systems (VDBMSs)}~\cite{xie2023briefvecdb}, such as Weaviate~\cite{weaviate}, Pinecone~\cite{pinecone}, and Qdrant~\cite{qdrant}, into the forefront as a foundational infrastructure for the data-intensive era~\cite{pan2024vdbmssurvey,taipalus2024vdbms,jin2024llmmeetsvecdb}. Unlike traditional databases optimized for tabular or document-based data, VDBMS specializes in storing, indexing, and querying vector embeddings—dense mathematical representations of unstructured data that power the core functionality of LLMs, recommendation engines, semantic search, and multimodal applications. VDBMS plays a pivotal role in enabling advanced LLM capabilities~\cite{jin2024llmmeetsvecdb,han2023vecdbsurvey,wang2024agenda}, such as \textbf{Retrieval-Augmented Generation (RAG)} systems~\cite{wang2024metmap}, \textbf{Long-Term Memory (LTM)} applications~\cite{zhao2024vision}, and LLM caching mechanisms~\cite{fu2023gptcache}. 
As LLM systems become increasingly sophisticated and data-hungry, VDBMS emerges as a critical technology for harnessing the full potential of these models and applications.

While embedding-based retrieval algorithms have been studied for over a decade in the academic community~\cite{wang2024crossmodalretrieval,dubey2022imageretrieval,guo2022semanticmodel}, the transition from theoretical frameworks to production-grade VDBMS marks a watershed moment for software testing. Over the past few years, this transition has spurred the rapid commercialization of VDBMS, with more than 40 specialized platforms now vying to optimize vector storage, real-time similarity search, and integration with LLM pipelines. \textbf{However, this explosive growth has not been matched by commensurate progress in rigorous software testing methodologies tailored for these emerging systems.} Unlike traditional databases optimized for structured data, VDBMS is purpose-built for storing, indexing, and querying vector embeddings. As such, they face unique testing challenges stemming from the high-dimensional nature of vector data~\cite{pan2024vdbmssurvey,xie2023briefvecdb,taipalus2024vdbms}, the approximate nature of vector search~\cite{indyk1998ann,huang2024ann}, and the need to support dynamic data scaling and hybrid query processing~\cite{pan2024vdbmssurvey,pan2024vdbms,guo2022manu}.

Current testing efforts for VDBMS are still in their infancy, primarily focusing on isolated performance benchmarks~\cite{vectordbbench,qdrantbenchmark,myscalebenchmark,martin2020annbenchmark} or simplified metamorphic testing cases~\cite{wang2024metmap} that address specific aspects like false vector matches in LLM-augmented generation systems. While these studies shed light on the critical impact of VDBMS defects on downstream applications, they represent initial steps, and \textbf{comprehensive testing methodology that holistically addresses the unique characteristics and challenges of VDBMS remains an open research gap.}
Lessons learned from the traditional \textbf{Database Management System (DBMS)} domain caution us that even after extensive research and rigorous testing, modern DBMS still exhibit systemic defects~\cite{tang2023dbmshiddenfailure,cui2024dbmstransaction,liang2024mozi,gao2023dbmsfuzz}. Similarly, the deep learning ecosystem, which shares similarities with VDBMS in terms of data-intensive workloads and complex computational pipelines, has encountered numerous reliability issues~\cite{chen2023dlbugs,islam2019dlbug,makkouk2022dlperformance}, potentially leading to crashes, hangs, build failures, and silent data corruption~\cite{tambon2024silentbug,wang2025sok}.
Given the pivotal role of VDBMS in enabling advanced LLM capabilities and the potentially far-reaching impact of failures, envisioning comprehensive testing tailored to VDBMS is imperative for the 2030 software engineering landscape.

However, testing VDBMS poses unique challenges that must be carefully considered:
\textbf{(1) The high-dimensional nature of vector data introduces complexities in test data generation, resource overhead, and performance trade-offs.}
High-dimensional vectors spanning across multiple storage pages cause memory explosion for fuzz testing tools~\cite{pan2024vdbms,pan2024vdbmssurvey,kersten2018everything}. Moreover, the $\mathcal{O}(d)$ time complexity of vector similarity computation, where $d$ denotes the vector dimension, drastically decreases testing speed compared to traditional $\mathcal{O}(1)$ attribute predicates~\cite{pan2024vdbms,pan2024vdbmssurvey}.
\textbf{(2) The fuzzy semantics in vector search make the oracle definition challenging.} ANN algorithms have non-deterministic $\epsilon$-error bounds~\cite{huang2024ann}, rendering traditional boolean assertions ineffective. 
\textbf{(3) Dynamic data scaling and hybrid query processing across data modalities demand testing strategies for evolving workloads and heterogeneous data types.} Data distribution shifts due to dynamic updates can cause indexing strategies to become imbalanced over time, often requiring full index rebuilds~\cite{pan2024vdbmssurvey}. Furthermore, hybrid queries involving predicate filtering before, during, or after vector similarity search across multimodal data necessitate diverse testing strategies.
\textbf{(4) The integration with complex LLM pipelines necessitates end-to-end testing to capture error propagation}, as small embedding perturbations can cascade into semantic errors after LLM decoding~\cite{wang2024metmap}, with error accumulation effects amplifying retrieval-stage issues~\cite{lancedbbug}. 

In light of these challenges, \textbf{this paper proposes the first roadmap for future research on VDBMS testing.} We begin by defining the common components and architecture of VDBMS, which typically consist of a vector storage engine, indexing subsystem, query processing pipeline, and client-side SDK. We then systematically analyze the bugs, vulnerabilities, and defects that have been discovered in existing VDBMS implementations, categorizing them into different components and symptoms. Drawing from these insights, we structure our investigation into three key aspects of VDBMS testing: test input generation, oracle definition, and test evaluation. For each of these aspects, we identify the unique challenges stemming from the intrinsic characteristics of VDBMS and propose a future research roadmap to address these challenges.

In summary, our contributions are detailed as follows:

\begin{itemize}[leftmargin=15pt]
    \item \textbf{Empirical Study of VDBMS Defects.} We conduct an empirical study of bugs across four major open-source VDBMS projects. Our analysis reveals a high prevalence of crash/hang bugs (23.1\%), with storage components severely impacted. Incorrect behavior bugs are most common (43.0\%), severely affecting query processing components. These findings highlight the need for tailored testing approaches to address critical defects in VDBMS.

    \item \textbf{Identification of Key Testing Challenges.} Drawing from the empirical study of defects in open-source VDBMS projects and the unique characteristics of VDBMS, we identify and articulate the key challenges in three fundamental aspects of VDBMS testing: test input generation, test oracle design, and test evaluation. These challenges are particularly critical in the VDBMS context due to the high-dimensional nature of vector data, the impact of dynamic data scaling on indexing strategies, and the complexity of vector operations and queries.

    \item \textbf{Proposal of the Research Roadmap.} Guided by the identified challenges, we propose the first comprehensive research roadmap that outlines future directions for addressing the critical aspects of VDBMS testing. Following this roadmap, the software testing community can contribute to the development of more reliable and trustworthy VDBMS, enabling the full potential of LLMs and data-intensive AI applications.
\end{itemize}

%% file: Chapters/2.vdbms.tex
\section{VDBMS}
\subsection{Architecture}
\begin{figure}[t]
\centering
\includegraphics[width=0.75\textwidth]{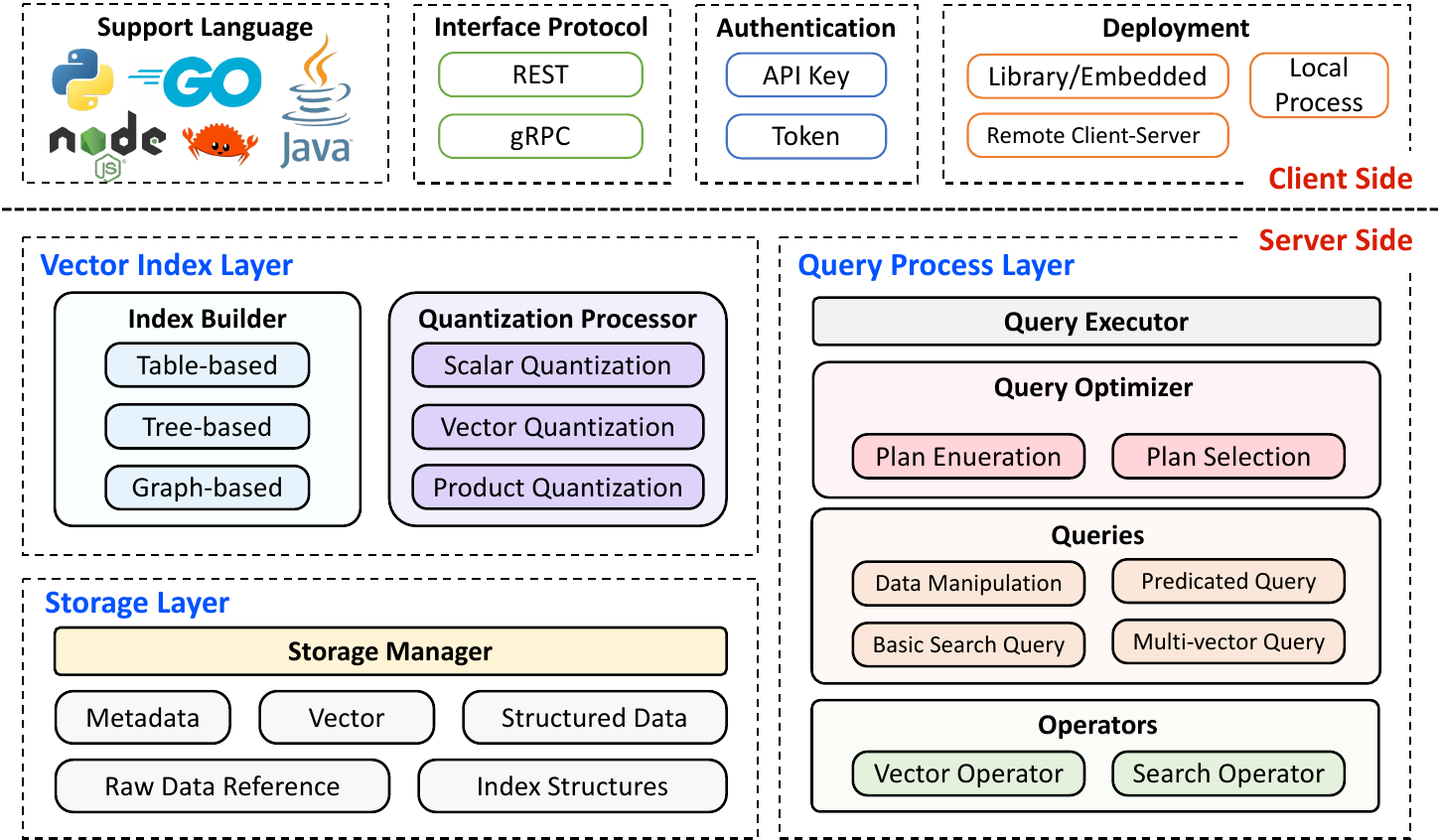}
\caption{Typical Architecture of VDBMSs.}
\label{fig:vdbms_arch}
\end{figure}

VDBMSs are specialized systems designed to efficiently store, index, and query high-dimensional vector embeddings. While the specific architectures can vary across different VDBMS implementations, they generally consist of several interconnected components, as shown in \autoref{fig:vdbms_arch}.

\noindent \textbf{Storage.} This layer handles the persistent storage of vector embeddings, associated metadata, raw data references, index structures, and structured data. The storage manager component oversees the storage and retrieval of these different data elements, often leveraging techniques like compression and partitioning for optimized space utilization.


\noindent \textbf{Vector Index.} The vector index layer enables efficient similarity search over vast vector collections through specialized indexing structures and quantization techniques tailored for high-dimensional data. The index builder constructs and maintains vector index structures based on table-based methods (e.g., LSH~\cite{indyk1998lsh}), tree-based methods (e.g., ANNOY~\cite{annoy}), and graph-based techniques (e.g., HNSW~\cite{malkov2020hnsw}). The quantization processor employs vector compression techniques, including scalar quantization, vector quantization, and product quantization.


\noindent \textbf{Query Process.} This layer is responsible for parsing, optimizing, and executing vector queries. It furnishes a set of operators tailored for data manipulation on vector tables. Search operators play a pivotal role in supporting similarity queries, enabling nearest neighbor retrieval or range searches.
Advanced query types cater to more intricate use cases. Predicated queries amalgamate vector conditions with structured data predicates. Multi-vector queries address scenarios where a single real-world entity is represented by multiple vectors, aggregating scores across these vectors. The query optimizer focuses primarily on predicated queries and systematically explores alternative execution strategies through plan enumeration. The query executor then implements the chosen plan through the coordinated use of distributed architectures and hardware acceleration.

\noindent \textbf{Client Side.} The client-side components of a VDBMS provide critical interfaces between end-users/applications and the underlying vector processing infrastructure.
VDBMS clients usually provide multi-language SDKs (Python, Go, Node.js, Java) with dual interface protocols REST APIs for metadata operations and gRPC~\cite{grpc} for high-throughput vector transfers. Security combines API key authentication with token-based authorization (JWT/OAuth2), while deployment flexibility spans embedded libraries, local standalone processes, and remote client-server architectures.

\subsection{Comparison of Existing VDBMS}
Modern VDBMSs exhibit significant architectural and functional diversity, as summarized in ~\autoref{tab:vdbms_comparison}. Our comparison focuses on key systems across critical dimensions that define modern vector data management capabilities, including system type, sub-type, implementation language, maximum supported vector dimension, indexing methods, and support for predicated queries. These diverse features and characteristics pose unique challenges for software testing and validation in the context of VDBMSs.

\begin{table}[t]
\centering
\fontsize{8}{11}\selectfont
\caption{Feature Comparison of 10 Representative VDBMSs.}
\label{tab:vdbms_comparison}
\begin{tabular}{lllcccc}
\toprule
\textbf{System} & \textbf{Type} & \textbf{Sub-type} & \textbf{Language} & \textbf{Max Dim} & \textbf{Index Methods} & \textbf{Pre. Query} \\
\midrule
Pinecone~\cite{pinecone} & Native & Vector & Rust & 20,000 & FreshDiskANN & \ding{52}  \\
Chroma~\cite{chroma} & Native & Vector & Python & Unlimited & HNSW & \ding{56}  \\
Weaviate~\cite{weaviate} & Native & Hybrid & Go & 65,535 & FLAT/FLAT-BQ/HNSW & \ding{52}  \\
Milvus~\cite{milvus} & Native & Hybrid & Go/C++ & Unlimited & FLAT/HNSW/ANNOY & \ding{52}  \\
Qdrant~\cite{qdrant} & Native & Hybrid & Rust & 65,536 & FLAT/HNSW/HNSW-PQ & \ding{52}  \\
Redis~\cite{redis} & Extended & NoSQL & C & Unlimited & FLAT/HNSW & \ding{52}  \\
MongoDB~\cite{mongodb} & Extended & NoSQL & C++/Java & 4,096 & FLAT/HNSW & \ding{52}  \\
Neo4j~\cite{neo4j} & Extended & NoSQL & Java/Scala & 2048 & HNSW & \ding{52}  \\
SingleStore~\cite{singlestoredb} & Extended & Relational & C++/Go & Unlimited & FLAT/IVF/HNSW & \ding{52}  \\
MyScale~\cite{myscale} & Extended & Relational & C++ & Unlimited & FLAT/MSTG/IVF & \ding{52}  \\
\bottomrule
\end{tabular}
\end{table}

Various systems are designed as native vector databases (e.g., Pinecone~\cite{pinecone}, Chroma~\cite{chroma}, Weaviate~\cite{weaviate}, Milvus~\cite{milvus}, Qdrant~\cite{qdrant}) or extend existing engines (e.g., Redis~\cite{redis}, MongoDB~\cite{mongodb}, SingleStoreDB~\cite{singlestoredb}, Neo4j~\cite{neo4j}, MyScale~\cite{myscale}) to incorporate vector capabilities. In terms of their data models, some systems are purely vector-oriented, while others adopt a hybrid approach combining vector and scalar data, necessitating comprehensive testing for handling different data types and their interactions. Turning to the implementation aspect, the languages used vary, requiring testing for language-specific issues and compatibility concerns across different technology stacks.
Another key factor influencing testing requirements is the maximum supported vector dimension, which ranges from specific limits to arbitrary high dimensions across different systems, demanding testing across diverse dimensionalities including boundary cases and extreme values. Additionally, the choice of indexing technique plays a crucial role in balancing search accuracy, performance, and memory efficiency. Indexing techniques like \textbf{Hierarchical Navigable Small World (HNSW)~\cite{malkov2020hnsw}}, \textbf{Inverted File Index (IVF)}, \textbf{Product Quantization (PQ)}, and \textbf{Approximate Nearest Neighbors Oh Yeah (ANNOY)~\cite{annoy}} are employed, with HNSW being particularly popular for large-scale similarity searches.
Finally, support for predicated queries, enabling filtered retrieval based on conditions, is an important feature in many vector database applications. Most of the listed systems support predicated queries, requiring rigorous testing of various predicate combinations and complexities to ensure accurate and efficient query processing.

%% file: Chapters/3.defects.tex
\section{Current State: Empirical Study of VDBMS Defects}
To understand the current state of software quality in VDBMSs, we conducted an initial empirical study on four prominent open-source projects in \autoref{tab:vdbms_comparison}: Milvus, Qdrant, Chroma, and Weaviate. 
To conduct a comprehensive analysis of bug-related issues and vulnerabilities in VDBMSs, we followed a rigorous process. As shown in \autoref{tab:vdbms_bug_analysis}, we first collected data on \textbf{Pull Requests (PRs)} from the repositories of four prominent open-source VDBMS projects. We then filtered these PRs using keywords related to bugs to obtain the subset of ``Bug PRs''. From this subset, we further filtered for closed and merged PRs, which resulted in a set of confirmed bug fixes. We manually analyzed and categorized based on the affected VDBMS components: storage, indexing, query processing, and client-side components. To ensure the accuracy and consistency of our analysis, we developed a set of criteria and guidelines for categorizing bugs based on their manifestation and impact on different VDBMS components. Two researchers independently analyzed and categorized a subset of the reported bugs. Any discrepancies or disagreements in the categorization were discussed and resolved through consensus.

\begin{table}[t]
\centering
\fontsize{8}{10}\selectfont
\caption{Analysis of Bug-Related Issues and Vulnerabilities in 4 Open Source VDBMSs.}
\label{tab:vdbms_bug_analysis}
\begin{tabular}{lrrrrrrc}
\toprule
\textbf{VDBMS} & \textbf{PRs} & \textbf{Bug PRs} & \multicolumn{4}{c}{\textbf{Manual Analyzed Bugs}} & \textbf{Vulnerabilities} \\
\cmidrule(lr){4-7}
& & & \textbf{Storage} & \textbf{Index} & \textbf{Query} & \textbf{Client} & \\
\midrule
Milvus & 9,576 & 414 & 26 & 5 & 42 & 35 & 13 \\
Qdrant & 1,035 & 35 & 10 & 0 & 3 & 13 & 8 \\
Chroma & 972 & 126 & 14 & 3 & 28 & 49 & 2 \\
Weaviate & 1,499 & 109 & 29 & 4 & 39 & 21 & 3 \\
\midrule
\textbf{Total} & 13,082 & 684 & 79 & 12 & 112 & 118 & 26 \\
\bottomrule
\end{tabular}
\end{table}

\begin{table}[t]
\centering
\fontsize{8}{10}\selectfont
\caption{Categorization of Bugs by Symptom and Component in Milvus/Qdrant/Chroma/Weaviate.}
\label{tab:bug_categorization}
\begin{tabular}{lccccr}
\toprule
\textbf{Symptom} & \textbf{Storage} & \textbf{Index} & \textbf{Query} & \textbf{Client} & \textbf{Total} \\
\midrule
Crash/Hang & 3/7/4/15 & 1/0/1/2 & 5/0/3/9 & 11/4/7/2 & 74~(23.1\%) \\
Build Failure & 0/0/0/0 & 0/0/0/0 & 0/0/0/0 & 0/5/20/6 & 31~(9.7\%) \\
Incorrect Behavior  & 3/1/10/10 & 0/0/2/2 & 17/2/24/28 & 16/2/13/8 & 138~(43.0\%) \\
Performance Issue & 6/2/0/4 & 1/0/0/0 & 5/0/1/2 & 7/1/1/0 & 30~(9.3\%) \\
Others & 14/0/0/0 & 3/0/0/0 & 15/1/0/0 & 1/1/8/5 & 48~(15.0\%) \\
\bottomrule
\end{tabular}
\end{table}

\autoref{tab:bug_categorization} presents the categorization of the manually analyzed bugs based on their symptoms and the affected VDBMS components. The data is ordered as Milvus/Qdrant/Chroma/Weaviate across each row. The table reveals several noteworthy patterns and highlights the prevalence of various types of bugs across different VDBMS components:

\begin{itemize}[leftmargin=15pt]
\item \textbf{Widespread Crash Bugs (23.1\%).} All four VDBMS systems exhibit a significant number of crash/hang bugs affecting storage, indexing, query processing, and client components. Such severe bugs can render the system completely inoperable, severely impacting its reliability and availability. Notably, the Weaviate system has the highest number of crash/hang bugs (15) related to its storage component. While the oracle design for detecting crashes is relatively straightforward (e.g., monitoring for process termination or unresponsiveness), generating effective test cases that trigger crash/hang bugs in VDBMSs poses unique challenges. 

\item \textbf{Prevalence of Incorrect Behavior (43.0\%).} Bugs that cause the system to exhibit incorrect or unexpected behavior, such as functional errors or incorrect output, are the most prevalent across all VDBMS components. Query processing components are particularly affected, with Milvus (17), Chroma (24), and Weaviate (28) exhibiting a high number of such bugs in query scenarios. Effectively detecting incorrect behavior in VDBMSs poses significant challenges for test oracles. In addition to validating the correctness of query results, test oracles must account for various aspects, including the accuracy of similarity scores, ranking order, and overall result quality. Furthermore, oracles for testing query components must handle diverse data distributions, high-dimensional vectors, and complex similarity functions. 

\item \textbf{Performance Degradation (9.3\%).} Performance-related bugs, which can lead to unacceptable system responsiveness or resource utilization, are consistently present across all VDBMSs. Detecting performance bugs in VDBMSs poses unique challenges. First, it requires generating complex test cases and workloads to construct and query large vector indexes, simulating real-world scenarios. Additionally, defining precise oracles for performance testing is more demanding, as it involves monitoring system metrics, resource consumption, and establishing acceptable performance thresholds. 

\item \textbf{Build and Integration Issues (9.7\%).} While build failures are relatively rare, client-side components in Chroma (20) and Weaviate (6) exhibit a significant number of such issues. These bugs can hinder the integration and deployment of these systems, potentially impacting their adoption and usability. VDBMS testing should include build and integration testing to ensure seamless deployment and usability, especially for client-side components that interface with various applications and environments.

\end{itemize}


%% file: Chapters/4.challenges.tex
\section{Challenges and Future Research Roadmap}
\begin{figure}[t]
\centering
\includegraphics[width=0.8\textwidth]{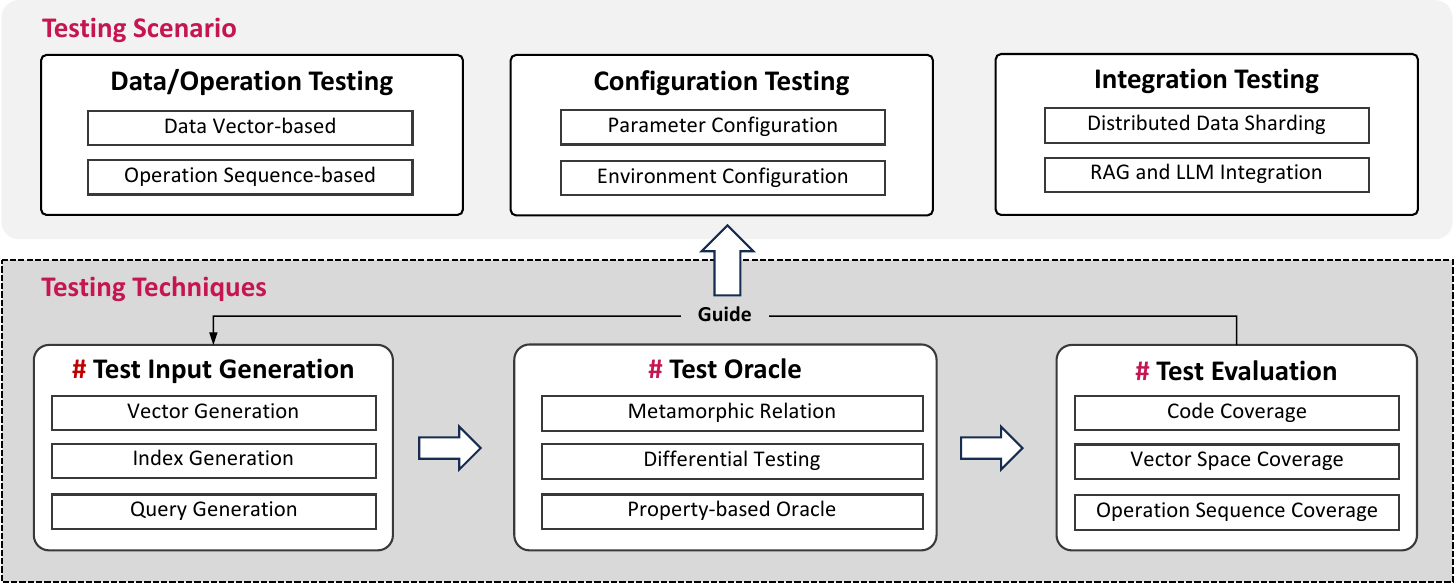}
\caption{Challenges and Future Research Roadmap for VDBMS Testing.}
\label{fig:roadmap}
\end{figure}

The empirical study of bug patterns and defects in prominent open-source VDBMSs reveals several critical challenges that must be addressed to improve the reliability of these emerging systems. As illustrated in \autoref{fig:roadmap}, we propose the first roadmap for future research on VDBMS testing. Specifically, we will discuss the key challenges and future research opportunities centered around three fundamental aspects: input generation, oracle definition, and test evaluation.

\noindent \textbf{Test Input Generation.} Test input generation for vector databases poses significant challenges due to the unique characteristics of vector data and the complexity of operations involved. The challenges and potential research opportunities can be categorized into the following aspects:

\begin{itemize}[leftmargin=15pt]

    \item \textbf{Vector Data Generation.} Generating representative input vector data is a critical challenge. High-dimensional vectors may exhibit characteristics like clustering, sparsity, and long-tailed distributions, making it difficult to generate test data that covers all potential scenarios. Research opportunities include developing techniques for generating high-dimensional, diverse, and representative vector data distributions. Additionally, the optimization techniques employed during vector storage and indexing processes may introduce precision errors or optimization traps, necessitating the generation of inputs that expose boundary conditions in floating-point operations, such as values close to zero, overflow, or underflow. Furthermore, VDBMSs support varying vector dimensions, requiring the generation of extreme value vectors or large-scale test inputs (e.g., 10 billion vectors) under resource constraints to test boundary conditions.

    
    \item \textbf{Index Generation.} Testing index construction in VDBMSs requires considering different index types (e.g., HNSW, IVF, LSH) and their parameter combinations (e.g., \texttt{M} and \texttt{efConstruction} for HNSW), which can lead to an exponential growth in the input space. Exhaustively testing all parameter combinations is impractical, necessitating the selection of representative test inputs. A potential research opportunity lies in exploring strategies for selecting representative test inputs for index construction, considering different index types, parameter combinations, and vector data characteristics. Additionally, scenarios like incremental indexing should be considered when generating targeted test cases.

    \item \textbf{Query Generation.} Test input generation should cover various query types, including not only basic data operations and search queries but also focusing on complex scenarios such as predicate queries and multi-vector queries. Predicate queries involve combinations of vector search and structured filtering conditions (e.g., $\text{WHERE } category=A \text{ AND } vector \approx [...] $), leading to an exponential input space due to the possible combinations of structured conditions (numeric ranges, string matching, boolean logic) and vector search conditions. A potential research opportunity is developing techniques for generating test inputs that cover such complex query scenarios. Multi-vector joint searches (e.g., $vector1 \approx [...] \text{ AND } vector2 \approx [...])$ require generating semantically related vector pairs (e.g., different modality features describing the same object), with diverse supported operations (e.g., set operations on vector collections, batch similarity computations). 
    Leveraging domain knowledge to guide the generation of semantically meaningful and relevant test inputs for vector databases is another potential research direction.
\end{itemize}

\noindent \textbf{Test Oracle.} Establishing effective test oracles for vector databases presents significant challenges due to the unique characteristics of vector data and the complexity of operations involved. While traditional test oracles for detecting crashes and hangs are still applicable, they account for a relatively small proportion of bugs (23.1\%) observed in VDBMSs. The majority of bugs (43.0\%) are related to incorrect behavior, necessitating the adaptation of test oracles to the specific features and nuances of vector databases.

\begin{itemize}[leftmargin=15pt]
    \item \textbf{Metamophic Relations (MRs)} One challenge lies in the design of MRs, which capture the expected relationships between inputs and outputs of a system under test. In the context of VDBMSs, the complexity of compound operations, such as transactional operations involving both insertions and queries, makes it difficult to formalize MRs that accurately capture the expected state consistency. Additionally, the inherent uncertainty introduced by high-dimensional data perturbations, where minor vector modifications (e.g., adding noise) may alter the similarity ordering, can render MRs ineffective. Future research should focus on developing systematic approaches for designing effective MRs that capture the intricate relationships between inputs, outputs, and system states in vector databases, while accounting for the complexities of compound operations and high-dimensional data perturbations.

    \item \textbf{Differential Tesing.} Differential testing, which compares the outputs of multiple systems or versions for the same inputs, may face challenges in interpreting result discrepancies. For instance, when performing approximate similarity searches, the difference between the top-k results returned by two systems may be ambiguous in terms of whether it falls within a reasonable error margin or indicates a legitimate bug. This ambiguity arises from factors like varying indexing parameters across systems, rather than inherent defects. Establishing principled techniques for interpreting and resolving ambiguities in result discrepancies, potentially leveraging domain knowledge or statistical methods to distinguish between legitimate bugs and expected approximations, is a crucial research direction.

    \item \textbf{Property-based Oracle.} Property-based testing, which involves checking if a system exhibits certain desirable properties, can be hindered by the difficulty in defining a comprehensive set of properties for high-dimensional vector data due to its complex mathematical nature. Furthermore, dynamic system states, such as index rebuilding, may lead to time-varying properties, necessitating the consideration of temporal dependencies in property definitions. Future research should explore techniques for eliciting and formalizing properties of vector databases, drawing from domain knowledge, data characteristics, and mathematical properties of high-dimensional spaces, while considering the dynamic nature of system states. 
\end{itemize}

\noindent \textbf{Test Evaluation.} Evaluating the effectiveness of testing VDBMSs poses unique challenges. While traditional code coverage metrics (e.g., line coverage, branch coverage) can partially reflect test effectiveness, they struggle to differentiate critical code segments and branches in the context of VDBMSs. API test coverage, which measures the coverage of specific important APIs, may not adequately capture the complexity of API parameter combinations and cross-API state dependencies. Future research should investigate techniques for measuring the coverage of API parameter combinations and cross-API state dependencies, potentially leveraging approaches such as operation sequence coverage or adapted variants.
Operation sequence coverage, which measures the coverage of sequences of operations, can be a more informative metric. However, it faces the challenge of prioritization bias, where randomly generated sequences may overly cover simple operations (e.g., single queries) while neglecting high-risk compound operations (e.g., ``insert → query → delete → index rebuild''). Additionally, validating long operation sequences can be computationally expensive. 
Due to the high-dimensional nature of vector data, evaluating vector space coverage may also be necessary. However, effective design and measurement of vector space coverage require careful consideration to ensure adequate coverage of sparse regions or cluster boundaries. 


%% file: Chapters/5.conclusion.tex
\section{Conclusion}
VDBMS have emerged as critical enablers for advanced LLM capabilities and data-intensive AI applications. However, the unique characteristics of VDBMS, including high-dimensional vector data, approximate vector search semantics, dynamic data scaling, and integration with complex LLM pipelines, pose significant challenges for software testing. In this paper, we have systematically analyzed these challenges and proposed the first comprehensive research roadmap for future work on VDBMS testing. By addressing the key aspects of test input generation, oracle definition, and test evaluation tailored to VDBMS, the software testing community can contribute to the development of more reliable and trustworthy VDBMS implementations.